\documentclass[prl,preprint,superscriptaddress,showpacs]{revtex4}

\usepackage{latexsym}
\usepackage{amsfonts}
\usepackage{epsfig}
\usepackage{amssymb}
\usepackage{mathrsfs}
\usepackage{amsmath}
\usepackage{graphicx}    

\begin{document}

\title{Effect of hydrogen bond cooperativity on the behavior of water}

\author{Kevin Stokely}
\affiliation{Center for Polymer Studies and Department of Physics,
   Boston University, Boston, Massachusetts 02215, USA}
\author{Marco G. Mazza}
\affiliation{Center for Polymer Studies and Department of Physics,
   Boston University, Boston, Massachusetts 02215, USA}
\author{H. Eugene Stanley}
\affiliation{Center for Polymer Studies and Department of Physics,
   Boston University, Boston, Massachusetts 02215, USA}
\author{Giancarlo Franzese}
\affiliation{Departament de Fisica Fonamental,
 Universitat de Barcelona, Diagonal 647, 08028 Barcelona, Spain}

\begin{abstract}

  Four scenarios have been proposed for the low--temperature phase behavior of liquid water, each predicting different thermodynamics.
  The physical mechanism which leads to each is debated.
  Moreover, it is still unclear which of the scenarios best
  describes water, as there is no definitive experimental test.
  Here we address both open issues within the framework of a microscopic cell model by performing a study combining mean field calculations
  and Monte Carlo simulations.
  We show that a common physical mechanism underlies each of the four scenarios, and
  that two key physical quantities determine which of the four
  scenarios describes water:  (i) the strength of the directional component of the hydrogen
  bond and (ii) the strength of the cooperative component of the hydrogen bond.
  The four scenarios may be mapped in the
  space of these two quantities.
  We argue that our conclusions are model-independent.
  Using estimates from experimental data 
for H bond properties
  the model predicts that the low-temperature phase diagram of water
  exhibits a liquid--liquid critical point at positive pressure.

\end{abstract}

\maketitle

Water's phase diagram is rich and complex: more than sixteen
crystalline phases \cite{Zheligovskaya}, and two or more glasses
\cite{Debenedetti-JPCM03, loerting, gruner} have been reported. The liquid state also displays interesting
behavior, such as the density maximum for 1 atm at $4^\circ$C.
The volume fluctuations $\left<(\delta V)^2\right>$, entropy
fluctuations $\left<(\delta S)^2\right>$, and cross-fluctuations between
volume and entropy $\left<\delta V \delta S\right>$,
proportional to the magnitude of isothermal
compressibility $K_T$,  isobaric specific heat $C_P$, and  isobaric
thermal expansivity $\alpha_P$, respectively, 
show anomalous increases in magnitude upon cooling \cite{angelldiv}.
Further, these quantities display an apparent
divergence for 1 atm at $-45^\circ$C \cite{angelldiv}, hinting at
interesting phase behavior in the supercooled region.

Microscopically, water's anomalous liquid behavior is understood as
resulting from the tendency of neighboring molecules to form hydrogen
(H) bonds upon cooling, with a decrease of local potential energy,
decrease of local entropy, and increase of local volume due to the
formation of local open structures of bonded molecules. 
Different models include these H-bond features, but depending on the assumptions and approximations of each model,
different conclusions are obtained for the
low--T phase behavior.
The relevant region of the bulk liquid state cannot be
probed experimentally, and none of the theories tested,
because crystallization of bulk water is
unavoidable below the homogeneous nucleation
temperature $T_H$ ($-38^\circ$C at 1~atm).


\section{Four scenarios for supercooled water}

Due to the difficulty of obtaining experimental evidence, theoretical and
numerical analyses are useful. Four separate scenarios for the
pressure--temperature ($P-T$) phase diagram have been proposed:

(I) The {\it stability limit} (SL) scenario~\cite{Speedy82}
hypothesizes that the superheated liquid-gas spinodal at negative
pressure re-enters the positive $P$ region below $T_H(P)$. In this view, the
liquid state is delimited by a single continuous locus, $P_s(T)$, bounding the
superheated, stretched and supercooled states. There is no reference to the phase into which the
liquid transforms when $P\to P_s(T)$. As the spinodal is approached, $K_T$, $C_P$, and $\alpha_P \to \infty$.
A thermodynamic consequence of the SL scenario is
that the intersection of the retracing spinodal with the liquid--vapor
coexistence line must be a critical point~\cite{Debenedetti-JPCM03}. The
presence of a such a critical point in the liquid--vapor transition, although
possible, is not confirmed by any experiment. This fact poses a serious
challenge to the SL scenario. 

(II) The \emph{liquid--liquid critical point} (LLCP) scenario
\cite{pses} hypothesizes a first--order phase transition line between
two liquids --- a low density liquid (LDL), and a high density liquid (HDL) ---
which terminates at a liquid--liquid
critical point $C'$. HDL is a dense liquid with a highly disordered structure,
whereas LDL has a lower density and locally tetrahedral  order. The experimentally
observable high density amorphous (HDA) and low density amorphous (LDA) solids correspond, in this scenario, to a structurally
arrested state of HDL and LDL respectively~\cite{StarrPRE99,BellissentJCP95}. Starting from
$C'$, the locus of maxima of the correlation length  $\xi$ (the Widom line) 
projects into the one--phase region~\cite{Xu-PNAS05}. Asymptotically close to
the critical point, response functions can be expressed in terms of $\xi$, hence,
these too will show maxima, e.g., as a function of $T$ upon isobaric cooling.
These maxima will diverge upon approaching $C'$. Furthermore, for
$P>P_{C'}$, the pressure of $C'$, the
response functions will diverge by approaching the spinodal converging to
$C'$. Specific models suggest~\cite{psgsa,pses} that $P_{C'}>0$, but
the possibility $P_{C'}<0$ has also been proposed~\cite{t}.

(III) The \emph{singularity--free} (SF) scenario~\cite{sdss}
hypothesizes that the low-$T$ anticorrelation between volume and
entropy is sufficient to cause the response functions to increase upon cooling
and display maxima at non--zero $T$, without reference to any singular behavior.
Specifically, Sastry \emph{et al.}~\cite{sdss} consider the temperature of maximum
density (TMD) line, where the
density has a maximum as a function of temperature, and 
prove a general thermodynamic theorem establishing the proportionality between the slope
of the  TMD, $(\partial P/\partial T)_{\mathrm{TMD}}$, and the temperature
derivative of $K_T$. Thus, since the TMD has negative slope in water, it follows that
$(\partial P/\partial T)_{\mathrm{TMD}}<0$, and therefore $K_T$ must increase
upon cooling, whether there exists a singularity or not.

(IV) The~\emph{critical--point free} (CPF) scenario \cite{a}
hypothesizes an order--disorder transition, with possibly a weak first--order
transition character, separating two liquid phases and extending to $P<0$ down
to the superheated limit of stability of liquid water. This scenario effectively
predicts a continuous locus of stability limit spanning the superheated,
stretched and supercooled state, because the spinodal associated with the
first--order transition will intersect the liquid--gas spinodal at negative
pressure. No critical point is present in this scenario. 

These four scenarios predict fundamentally different behavior, though
each has been rationalized as a consequence of the same microscopic interaction: the H bond.
A question that naturally arises is whether the
macroscopic thermodynamic descriptions are in fact connected in some
way.  Previous works have attempted to uncover relations between several
of the scenarios, for example between (I) and (II) \cite{psgsa,bds}
or (II) and (III) \cite{tdst,tanaka2000}.
Here we offer a relation linking all four scenarios showing
that (a) all four can be included in one general scheme, (b) the
balance between the energies of two components of the H bond
interaction determines which scenario is valid.
Morevover, we argue that current values for these energies support the LLCP scenario.

\section{Cooperative cell model of water}



We analyze a microscopic model~\cite{fs} of water in which the
fluid is divided into $N$ cells with
nearest neighbor (n.n.) interactions.
The division is such that each cell is in contact
with four n.n., mimicking the first shell of liquid water.
The case of a shared H bond, due to more than four molecules in the first shell, is assimilated with the case in which a H bond is broken, since the interaction energy of a shared bond is less than half the energy of a single H bond \cite{sfs,ek}.


The goal of the model is to represent, microscopically, the
essential features of the interaction among water molecules, while
being able to qualitatively understand the importance of each of
these features.  To this end the interaction among cells is separated
into four distinct components.

The first component of the interaction is due to the short--range repulsion of the electron clouds.
This is incorporated into the model by assigning to each cell $i\in [1,N]$ (a) a volume $v_i\geq v_0$,
where $v_0$ is the exclusion volume per molecule, and (b) a maximum of one molecule.

The second component includes all the isotropic long--range attractive interactions,
such as the instantaneous induced dipole-dipole (London) interactions between the electron clouds of different molecules
or the isotropic part of the hydrogen bond~\cite{hbond}.
We refer to this component as the van der Waals attractive interaction,
keeping in mind, however, that this component includes not only the (weak) London dispersion interaction,
but also the (stronger) isotropic interaction of the hydrogen bond. 
The overall sum of the isotropic ---attractive and repulsive---
interactions can be represented in different ways.
The one we adopt in a mean field (MF) treatment is
\begin{equation}
\label{vdw}
\mathscr{H}_{\rm o} \equiv - \epsilon \sum_{\langle i,j \rangle} n_i n_j ,
\end{equation}
\noindent
where if $v_0/v_i\leq 0.5$ we set the index $n_i=0$, and if $0.5<v_0/v_i \leq 1$ we set $n_i=1$,
hence $n_i=0$ if the density in the cell is gas-like, and $n_i=1$ if the density in the cell is liquid-like; 
$\epsilon>0$ is the characteristic energy of the attraction and the sum is over all n.n. pairs $\langle i,j \rangle$.

The characteristic feature of H$_2$O is its ability to form H bonds between neighboring molecules.
This interaction has a strong directional component due to the dipole-dipole interaction between the highly
concentrated positive charge on each H and each of the two excess negative charges concentrated on the O of another water molecule.
Accordingly, the third component incorporated here is this orientational--dependent interaction,
which includes the covalent component of the bond~\cite{hbond0}.
To account for the orientational degrees of freedom of each water molecule, we assign to each cell $i$
four bond variables $\sigma_{ij}=1,...,q$ (one for each n.n. cell $j$),
representing the orientation of molecule $i$ with respect to molecule $j$.
We choose the parameter $q$ by selecting 30$^o$ as the maximum deviation from a linear bond, i.e. $q\equiv 180^o/30^o=6$,
hence every molecule has $q^4=6^4\equiv 1296$ possible orientations.
(The effect of choosing a different value for $q$ has been analyzed in~\cite{fs-JPCM07}.) 
We say that a bond is formed between cells $i$ and $j$ if $\sigma_{ij}=\sigma_{ji}$.

Experiments show that formation of the H bonds leads to an open ---locally tetrahedral---
structure that induces an increase of volume per molecule~\cite{Debenedetti-JPCM03,soperricci}.
This effect is incorporated in the model by considering the total volume to be given as
\begin{equation}
\label{vol}
V \equiv N v_0 + N_{\rm HB} v_{\rm HB},
\end{equation}
\noindent
where
\begin{equation}
N_{\rm HB}\equiv
\sum_{\langle i,j \rangle} n_i n_j  \delta_{\sigma_{ij},\sigma_{ji}}
\end{equation}
\noindent
is the total number of H bonds, $\delta_{a,b}=1$ if $a=b$, 
$\delta_{a,b}=0$ if $a\neq b$, and $v_{\rm HB}$ is the volume increase per H bond~\cite{sdss}.
Bond formation also leads to a decrease in the local potential energy, hence we add to the
Hamiltonian in Eq.~(\ref{vdw}) the term
\begin{equation}
\label{hb}
\mathscr{H}_{\rm HB} \equiv - J N_{\rm HB} ,
\end{equation}
\noindent
where $J>0$ is the characteristic energy of the directional component of the H bond.


Another key experimental fact is that at low $T$ the O--O--O angle
distribution in water becomes sharper around the tetrahedral value~\cite{ricci},
suggesting an interaction that induces a cooperative behavior among bonds.
For water, four--body and higher order interactions seem to be negligible
with respect to the three--body term~\cite{Pedulla-JCP96,Skinner-JPCB08}.
Hence, the fourth component to the interaction potential is the many--body effect due to H bonds~\cite{hbond1,hbond2,hbcoop},
which minimizes the energy when the H bonds of nearby molecules assume a tetrahedral orientation.
This is accomplished by further adding to the Hamiltonian in Eqs.~(\ref{vdw}) and (\ref{hb})
the term
\begin{equation}
\mathscr{H}_{\rm coop} = -J_\sigma \sum_i n_i \sum_{(k,\ell)_i} \delta_{\sigma_{ik},\sigma_{i\ell}} ,
\end{equation}
\noindent
where $J_\sigma>0$ is the characteristic energy of the cooperative component of the H bond, and $(k,\ell)_i$ indicates one of the six different pairs of the
four bond variables of molecule $i$.  This interaction introduces a
cooperative behavior among bonds, which may be fine tuned by changing
$J_\sigma$.  Choosing $J_\sigma=0$ leads to H bonds which form
independent of neighboring bonds~\cite{sdss}, while $J_\sigma \rightarrow \infty$
leads to fully dependent bonds~\cite{sss}.
The total Hamiltonian is now given by
\begin{equation}\label{ham}
\mathscr{H} =\mathscr{H}_{\rm o}+\mathscr{H}_{\rm HB}+\mathscr{H}_{\rm coop} .
\end{equation}

This model is studied using both MF analysis and Monte
Carlo (MC) simulations~\cite{fms,kfs-PRL08,fs-JPCM07,fs-PhysA02,kfs-JPCM08}.
Details of the MF and MC techniques are available elsewhere~\cite{fs-JPCM07,mazza}.
In the following we adopt $\tilde{J}\equiv J/\epsilon$,
$\tilde{J_\sigma}\equiv J_\sigma/\epsilon$ and $v_{\rm HB}=2v_0$.

\section{Mean-field results}

\begin{figure}
\includegraphics[scale=0.25]{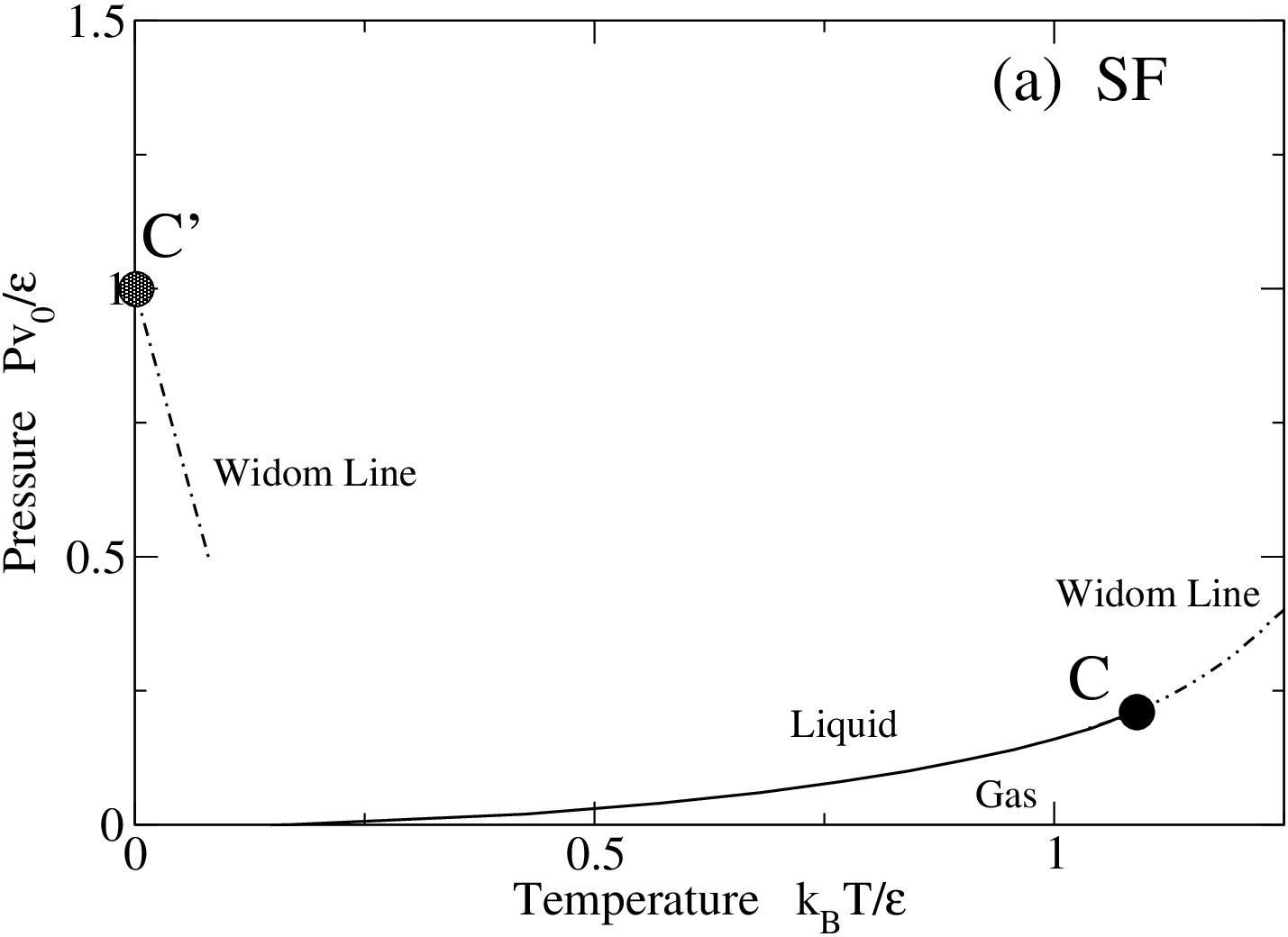}
\includegraphics[scale=0.25]{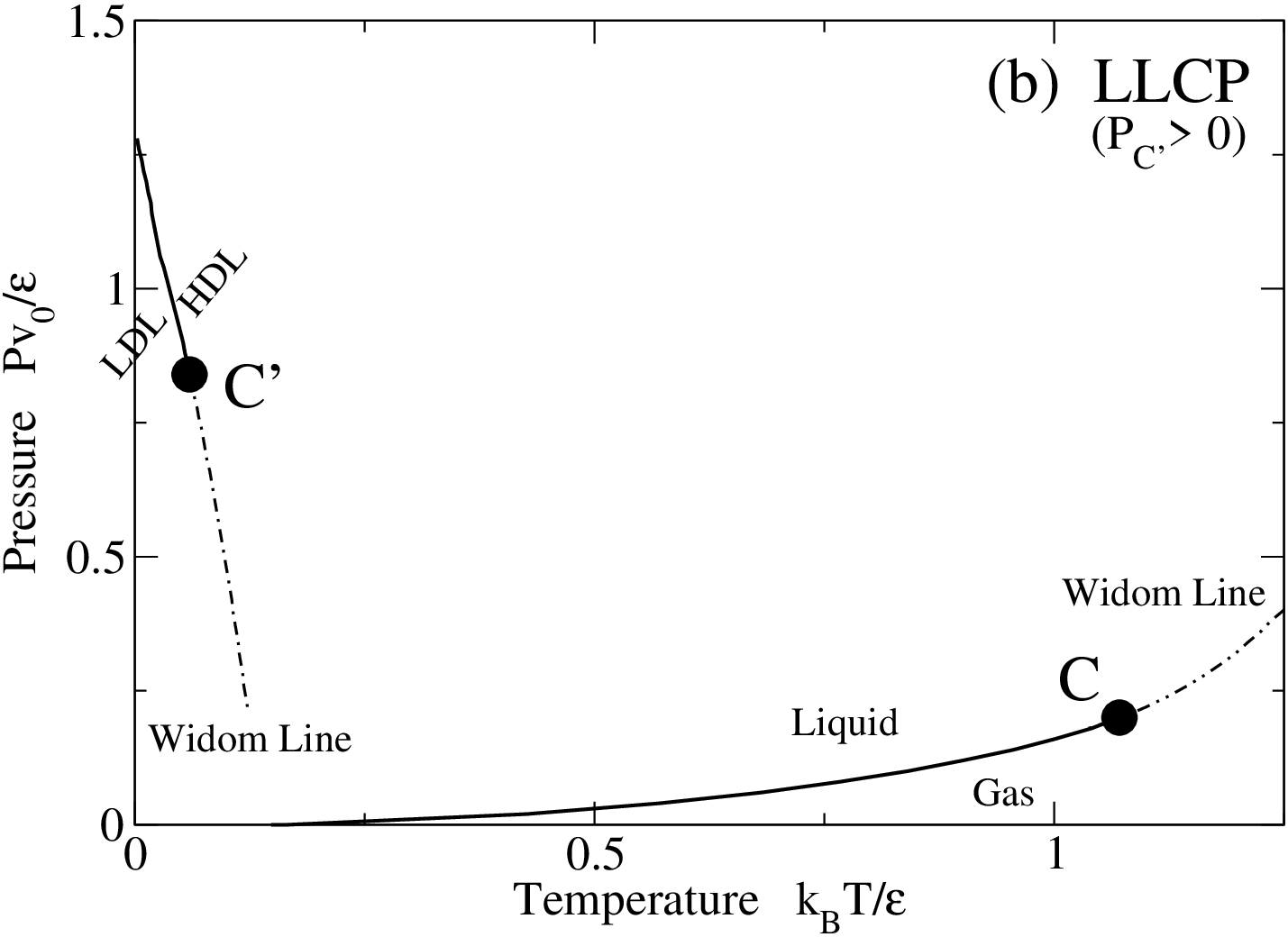}
\includegraphics[scale=0.25]{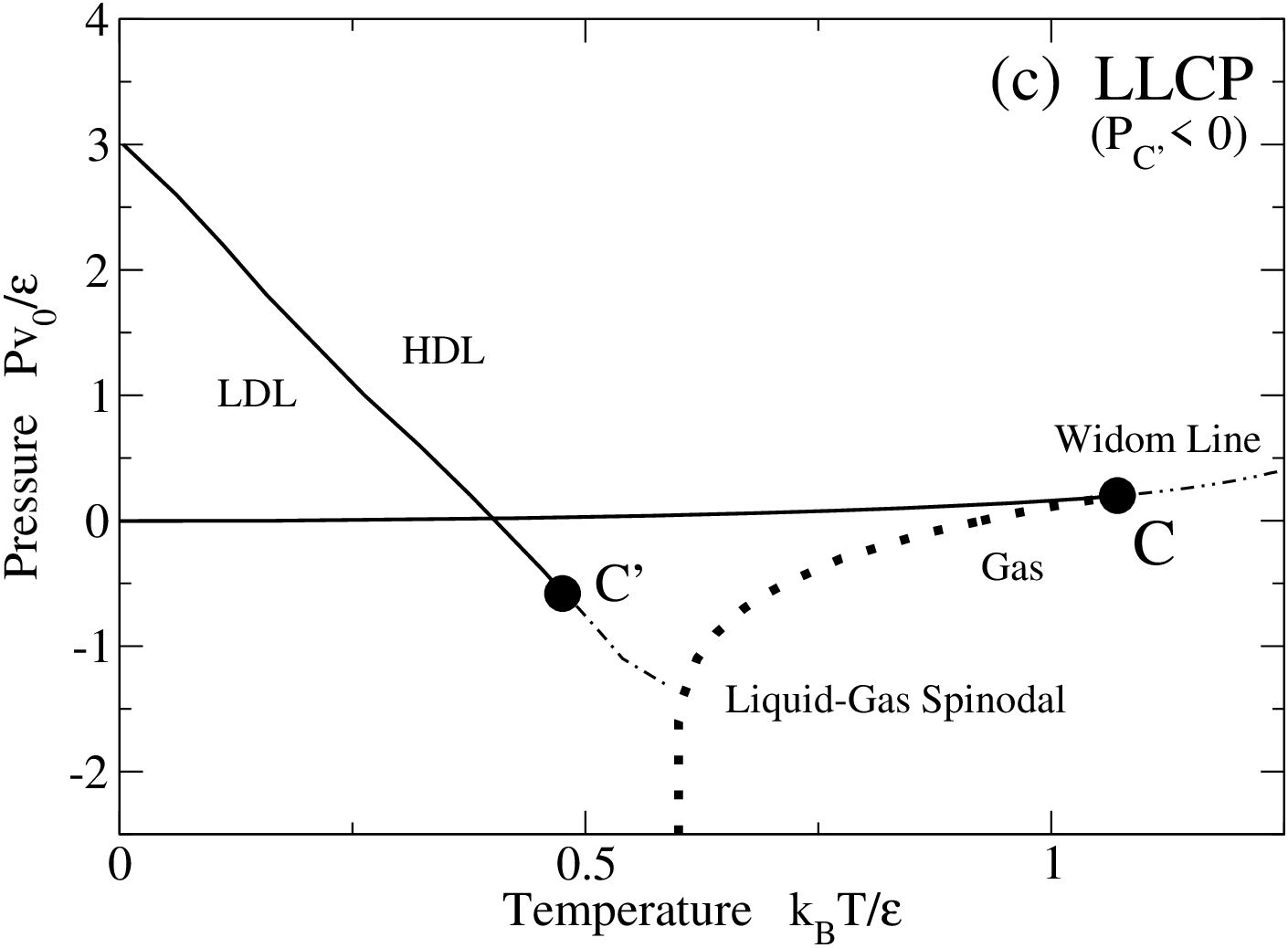}
\includegraphics[scale=0.25]{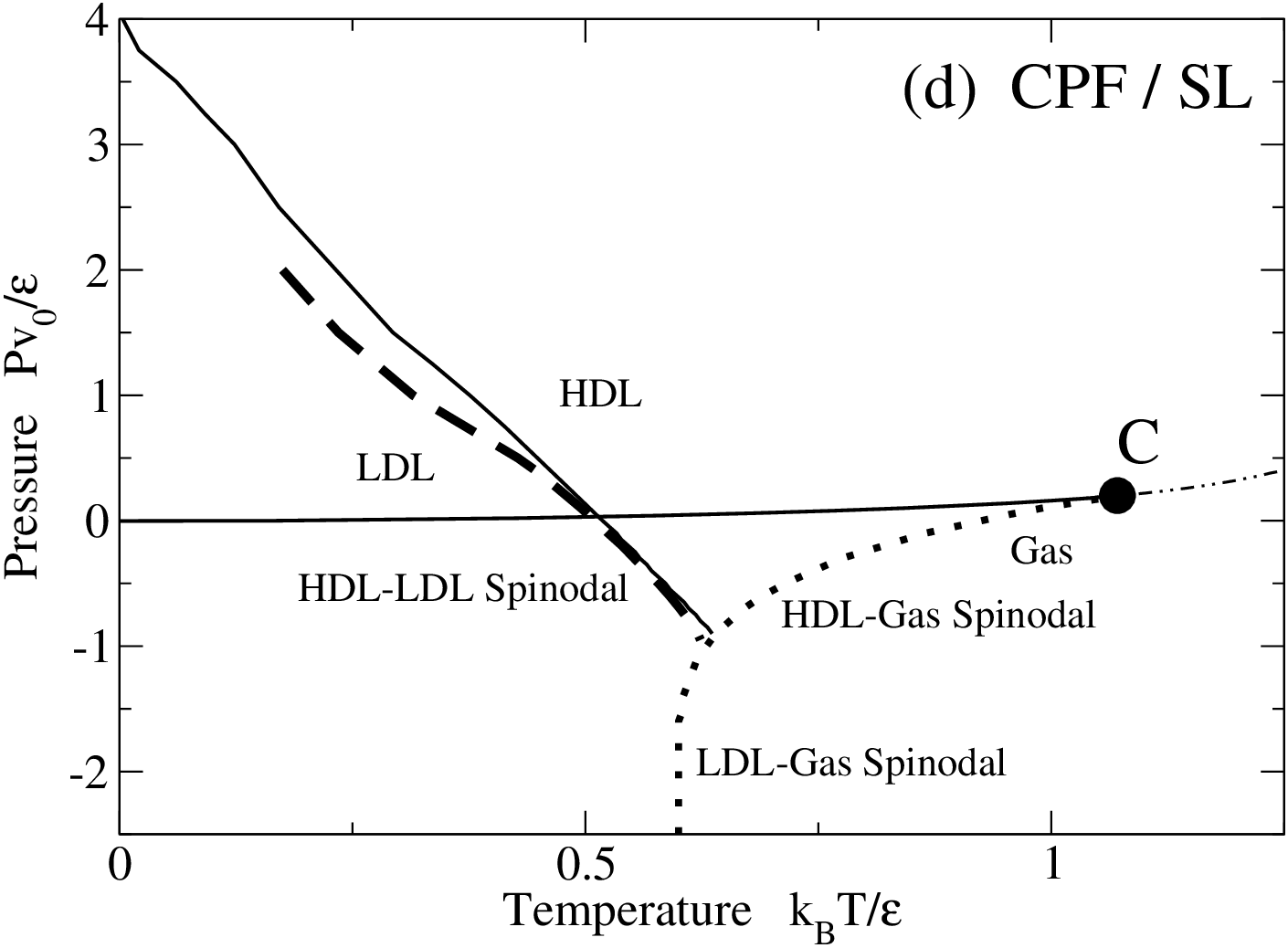}
\caption{
Phase diagram predicted from MF calculations for the cell model with
    fixed H bond strength ($\tilde{J}=0.5$), fixed H bond volume increase ($v_{\rm HB}/v_0=0.5$),
    and different values of the H bond cooperativity strength $\tilde{J}_\sigma$.
(a) \emph{Singularity-free} scenario ($\tilde{J}_\sigma=0$).
    At high $T$, liquid (L) and gas (G) phases are separated by a first order
    transition line (thick line) ending at a critical point $C$,
    from which a L--G Widom line (double--dot--dashed line) emanates.
    In the liquid phase, maxima of  $K_T$ and $\alpha_P$ converge to a locus (dot--dashed line).
    At $C'$ both $K_T$ and $\alpha_P$ and have diverging maxima. The locus of the
   maxima is related to the L-L Widom line for $T_{C'}\rightarrow 0$ (see text).
(b) \emph{Liquid--liquid critical point} scenario with positive critical pressure (for $\tilde{J}_\sigma = 0.05$).
    At low $T$ and high $P$, a high density liquid (HDL) and a low density liquid (LDL) are separated by a
    first order transition line (thick line with HDL/LDL labeled) ending in a
    critical point $C'$, from which the L-L Widom line (dot--dashed line) emanates.
    Other symbols are as in the previous panel.
(c) \emph{Liquid--liquid critical point} scenario with negative critical pressure (for $\tilde{J}_\sigma = 0.35$).
    Here the L-L Widom line (dot--dashed line) is shown intersecting the L-G spinodal (dotted line).
    Other symbols are as in the previous panel.
(d) \emph{Critical--point free} scenario ($\tilde{J}_\sigma=0.5$).
    The HDL--LDL coexistence line extends to the superheated liquid region at $P<0$,
    reaching with the liquid spinodal (dotted line). The stability limit (SL) of
    water at ambient conditions (HDL) is delimited by the superheated
    liquid--to--gas spinodal and the supercooled HDL--to--LDL spinodal (dashed line),
    giving a re-entrant behavior as hypothesized in the SL scenario.
    Other symbols are as in the previous panels.
In all panels, $k_B$ is the Boltzmann constant.
}
\label{1}
\end{figure}

Three qualitatively different phase diagrams are found,
dependent on the strengths of the H bond energy parameters, $\tilde{J}$ and $\tilde{J_\sigma}$ (Fig.~\ref{1}).

When $\tilde{J_\sigma}=0$ the model coincides with that proposed in \cite{sdss},
which gives rise to the SF scenario (Fig.~\ref{1}a).
For $0 < \tilde{J_\sigma} \leq \tilde{J}/2$ the model displays a liquid--liquid transition
ending in a LLCP at $P_{C'}\geq 0$ (Fig.~\ref{1}b) \cite{fms}. 
For  $\tilde{J}/2 < \tilde{J_\sigma}/ < a+b \tilde{J}$, where $a=0.30\pm001$ and
$b=0.36\pm0.01$ are fitting parameters, a LLCP occurs at $P_{C'}<0$ (Fig.~\ref{1}c).
For $\tilde{J_\sigma} \geq a+b \tilde{J}$, a liquid-liquid transition
with no critical point is found, consistent with the CPF scenario (Fig.~\ref{1}d).
In Fig.~\ref{2} we summarize these results in the $\tilde{J}$ vs. $\tilde{J_\sigma}$
parameter space.

\subsection{Limiting behavior between the four cases}
 In the following we discuss how, by tuning $\tilde{J}$ and
 $\tilde{J_\sigma}$, we can pass from one scenario to another in a
continuous way.

(i) By beginning with the LLCP scenario, and studying the limit
$\tilde{J_\sigma} \rightarrow 0$, we find $T_{C'} \rightarrow 0$.
Moreover, we find that $K_T$ and $\alpha_P$ diverge as $|T-T_{C'}|^{-1}$ for
any value of $J_\sigma$, including $J_\sigma \rightarrow 0$ and $T_{C'} \rightarrow 0$.
Further, we find for the entropy $S$ that, for any value of $J_\sigma$,
$(\partial S/\partial T)_P \propto |T-T_{C'}|^{-1}$.
Hence $C_P\equiv T(\partial S/\partial T)_P$ diverges as $|T-T_{C'}|^{-1}$ when $T_{C'}>0$.
When $T_{C'}=0$ ($\tilde{J_\sigma}=0$), $C_P$ is constant, as in the SF scenario \cite{rebelo}. 
Therefore, the SF scenario coincides with the LLCP scenario in the limiting case of $T_{C'} \rightarrow 0$,
for $\tilde{J_\sigma} \rightarrow 0$ (Fig. \ref{1}a).

(ii) Again, beginning with the LLCP scenario, and increasing $\tilde{J_\sigma}$ while keeping other parameters constant,
we observe that $C'$ moves to larger $T$ and lower $P$, with $P_{C'}<0$ for $\tilde{J_\sigma} \geq \tilde{J}/2$ (Fig.\ref{1}c).

(iii) With further increase of $\tilde{J_\sigma}$, $C'$ approaches, and eventually reaches, the liquid--gas spinodal.
For larger values of $\tilde{J_\sigma}$ only the liquid--liquid transition remains, which is precisely the CPF scenario \cite{a} (Fig.~\ref{1}d).
Hence the CPF scenario differs from the LLCP scenario only in that $C'$ is now
inaccessible, lying beyond the region of liquid states.
The same result may be obtained by decreasing $\tilde{J}$, while fixing $\tilde{J_\sigma}$ and other parameters.
Here a decrease of $J$ moves $C'$ to lower pressure, i.e. towards the liquid spinodal, while the entire liquid-liquid
phenomena moves to succesively lower temperature.
In all cases, the location of $C'$ varies continuously with variation of $\tilde{J}$ and $\tilde{J_\sigma}$.

(iv) In the case of the CPF scenario, we find that the superheated liquid-gas spinodal
merges with the supercooled liquid-liquid spinodal, as in Ref. \cite{psgsa}.
This gives rise to a liquid spinodal which retraces in the $P$--$T$ plane.
This feature resembles the main characteristic of the SL scenario,
where the high-$T$ liquid has a limit of stability at $P<0$ that retraces toward $P>0$ at low $T$.
Here this retracing locus is formed by two spinodal lines, with different signs of the slope, that merge at $P<0$.
Therefore, in the framework of the present model, the CPF scenario and the SL scenario coincide,
corresponding to the case in which the cooperative behavior is very strong.

\subsection{Linearity of the lines separating one
  scenario from another in $\tilde{J}$--$\tilde{J_\sigma}$ plane}
For the cell model, we can derive 
\begin{equation}
T_{C'}=\tilde{J}_\sigma/\alpha+\mathscr{O}(\tilde{J}_\sigma^2)
\label{TC'} 
\end{equation}
and 
\begin{equation}
P_{C'}=(\tilde{J}^*/v_{\rm HB})+\beta T_{C'}+\mathscr{O}(T_{C'}^2).
\label{PC'}
\end{equation}
Here $\alpha>0$ and $\beta<0$ are constants and, in the MF context, $\tilde{J}^*\equiv \tilde{J}+3 \tilde{J}_\sigma$. 
Symbols $\mathscr{O}(X^2)$, where $X$ is  $\tilde{J}_\sigma$ or
$T_{C'}$, represent terms of order $X^2$ or higher, that are negligible when $X\ll1$.
Our MF results confirm the relations in Eq.~(\ref{TC'}) and (\ref{PC'}), 
with $\alpha\simeq 0.74 k_B/\epsilon$ and $\beta\simeq -7.4
k_B/v_0$, with negligible $\mathscr{O}(X^2)$ terms.

\begin{figure}
\begin{center}
\includegraphics[scale=0.58]{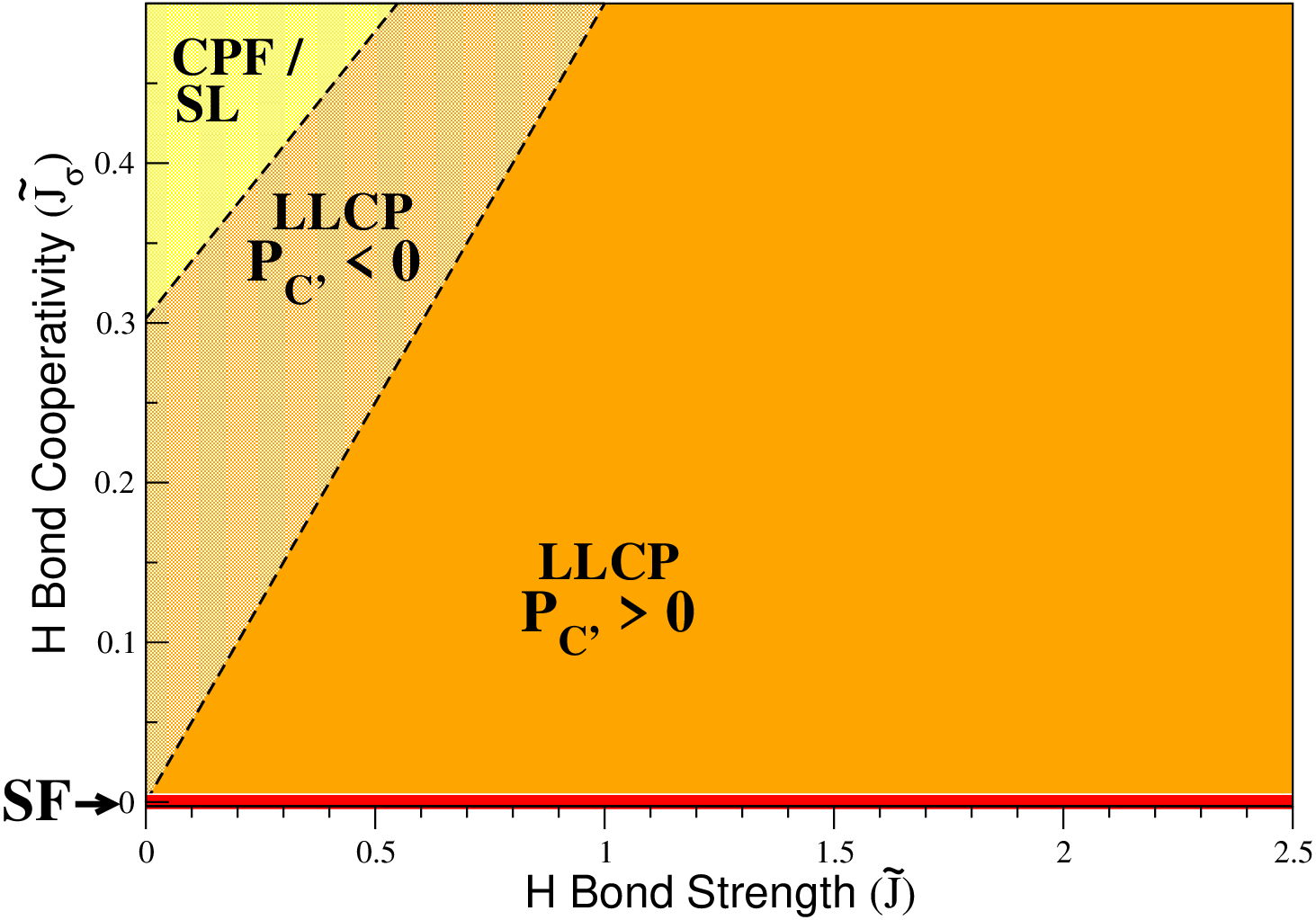}
\end{center}
\caption{
Possible scenarios for water for different values of H bond energies $\tilde{J}$,
  (directional component) and $\tilde{J}_\sigma$ (cooperative component),
  both in units of the van der Waals energy $\epsilon$, obtained
  from MF calculations.
  The ratio $v_{\rm HB}/v_0$ is kept constant.
(i) If $\tilde{J}_\sigma=0$ (red line along x--axis), the singularity free (SF) scenario
  is realized, independent of $\tilde{J}$.
(ii) For large enough $\tilde{J}_\sigma$, water would possess a first--order liquid--liquid phase
  transition line terminating at the liquid--gas spinodal---the
  critical point free (CPF) scenario; the liquid spinodal would
  retrace at negative pressure, as in the stability limit (SL)
  scenario (yellow region in top left).
(iii) For other combinations of $\tilde{J}$ and
  $\tilde{J}_\sigma$, water would be described by the liquid--liquid critical
  point (LLCP) scenario. For larger $\tilde{J}_\sigma$, the LLCP is at negative
  pressure (ochre region between dashed lines). For smaller $\tilde{J}_\sigma$, the LLCP is at
  positive pressure (orange region in bottom right). Dashed lines separating the three
  different regions correspond to mean field results of the
  microscopic cell model. Equations for the lines are
  $\tilde{J}_\sigma=\tilde{J}/2$ and
  $\tilde{J}_\sigma=a+b\tilde{J}$, with $a\simeq
  0.3$ and $b\simeq 0.36$.  The $P-T$ phase diagram evolves continuously
  as $\tilde{J}$ and $\tilde{J}_\sigma$ change.
}
\label{2}
\end{figure}

Therefore, we can rewrite the above relations
as
$\tilde{J}-P_{C'}v_{\rm HB}/\epsilon=-(3+\beta
v_{\rm HB}/\alpha)\tilde{J}_\sigma\equiv 2 \tilde{J}_\sigma$, when $v_{\rm HB}=2v_0$.
As a consequence, for the case $P_{C'}=0$, we find $\tilde{J}_\sigma=\tilde{J}/2$, which
is exactly what we find numerically in Fig.~\ref{2} along 
the line separating the LLCP scenario with $P_{C'}>0$ (valid for
$\tilde{J}_\sigma<\tilde{J}/2$) and the LLCP scenario with $P_{C'}<0$ (valid for
$\tilde{J}_\sigma>\tilde{J}/2$).

It is possible to show that Eq.~(\ref{PC'})
can be generalized to
$P_{LL}=(\tilde{J}^*/v_{\rm HB})+\beta T_{LL}+\mathscr{O}(T_{LL}^2)$, where $T_{LL}$ and
$P_{LL}$ are the $T$ and $P$ along the liquid--liquid transition line.
Our MF results are in good agreement with this prediction.

We can estimate the equation of
the line separating the LLCP scenario with $P_{C'}<0$ and the CPF/SL
scenario in the $\tilde{J}$--$\tilde{J}_\sigma$ plane,
by using the Eq.~(\ref{PC'}), together with the
equation for the liquid--gas spinodal.
In particular, we adopt a parametric fit, in terms of the parameter $\tilde{J}$, of the spinodal
pressure with respect to the spinodal temperature, and we evaluate the line separating the
LLCP
and CPF/SL scenarios
for $\tilde{J}\rightarrow 0$  when $C'$ is on the spinodal. 
From this approximate approach, we derive that 
$\tilde{J}_\sigma=\tilde{J}_\sigma^0+\gamma\tilde{J}$, with
$\tilde{J}_\sigma^0\simeq 0.2$, of the same
order of magnitude of the fitting parameter $a\simeq 0.30$ in Fig.~\ref{2}.  
Yet, $\gamma\neq 0.36$, the value of $b$ in Fig.~\ref{2}, as a
consequence of the strong approximations made.

\section{Monte Carlo results}

To test the validity of our MF calculations, we perform MC simulations in the $NPT$
ensemble %
\cite{mazza}. To this end,

(i) we consider that the total volume is $V\equiv V_{MC}+N_{\rm HB}v_{\rm HB}$, where
$V_{MC}\geqslant N v_0$ is a dynamical continuous variable;

(ii) we assume that the system is homogeneous with all the variables $n_i$
set to 1; with this assumption the gas state occurs when 
$\rho\equiv N/V< 0.5/v_0$;

(iii) we replace the isotropic repulsive and attractive terms of
the Hamiltonian in Eq.~(\ref{ham}) with a Lennard--Jones potential, more suitable for continuous distances $r$ between particles,  with
attractive energy $\epsilon>0$ plus a hard--core repulsion at distance $r_0$
\begin{equation}
U_W(r)\equiv\begin{cases}
\infty & \text{if $r\leqslant r_0$,}\\
\epsilon\left[\left(\frac{r_0}{r}\right)^{12}-
     \left(\frac{r_0}{r}\right)^6\right] & \text{if $r>r_0$}.\\
\end{cases}
\label{LJ}
\end{equation}


\noindent
Here $r_0\equiv(v_0)^{1/d}$ and $d$ is the system dimension
\cite{fms} (the hard--core repulsion reduces the
computational cost and does not change the phase diagram); the distance between two
n.n. molecules is $(V/N)^{1/d}$, and the distance $r$ between two
generic molecules is the Cartesian distance between the centers of the
cells in which they are enclosed.

(iv) We consider the system in $d=2$ dimensions. While the MF results
are valid for any dimension so long as the number of n.n. molecules is four,
the MC results hold for a system with 
coordination number four and two dimensions. Since the results in the
two cases are qualitatively comparable, we do not expect a strong
dependence of the phase diagram on dimension.

\begin{figure}
\begin{center}
\includegraphics[scale=0.3]{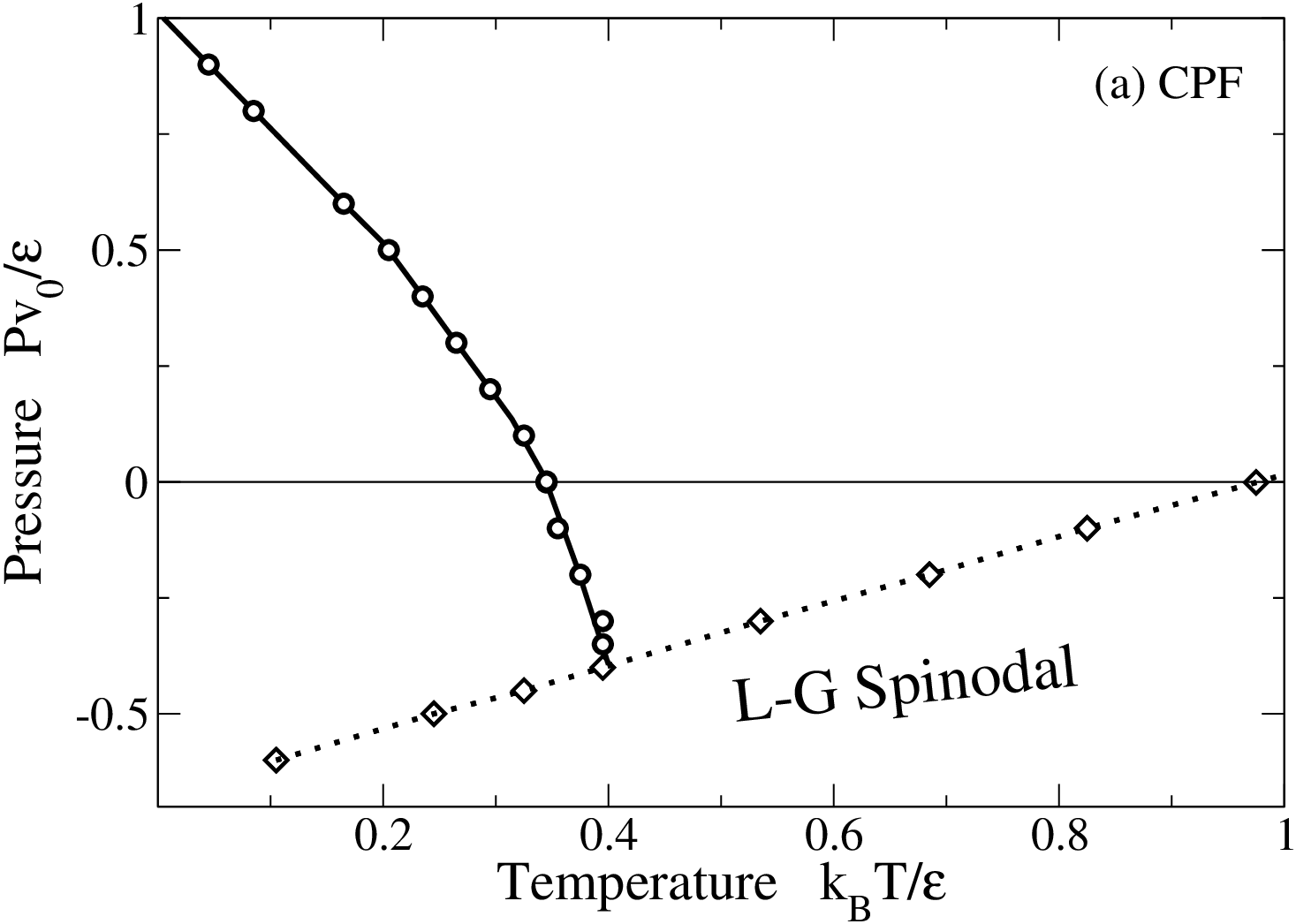}
\includegraphics[scale=0.3]{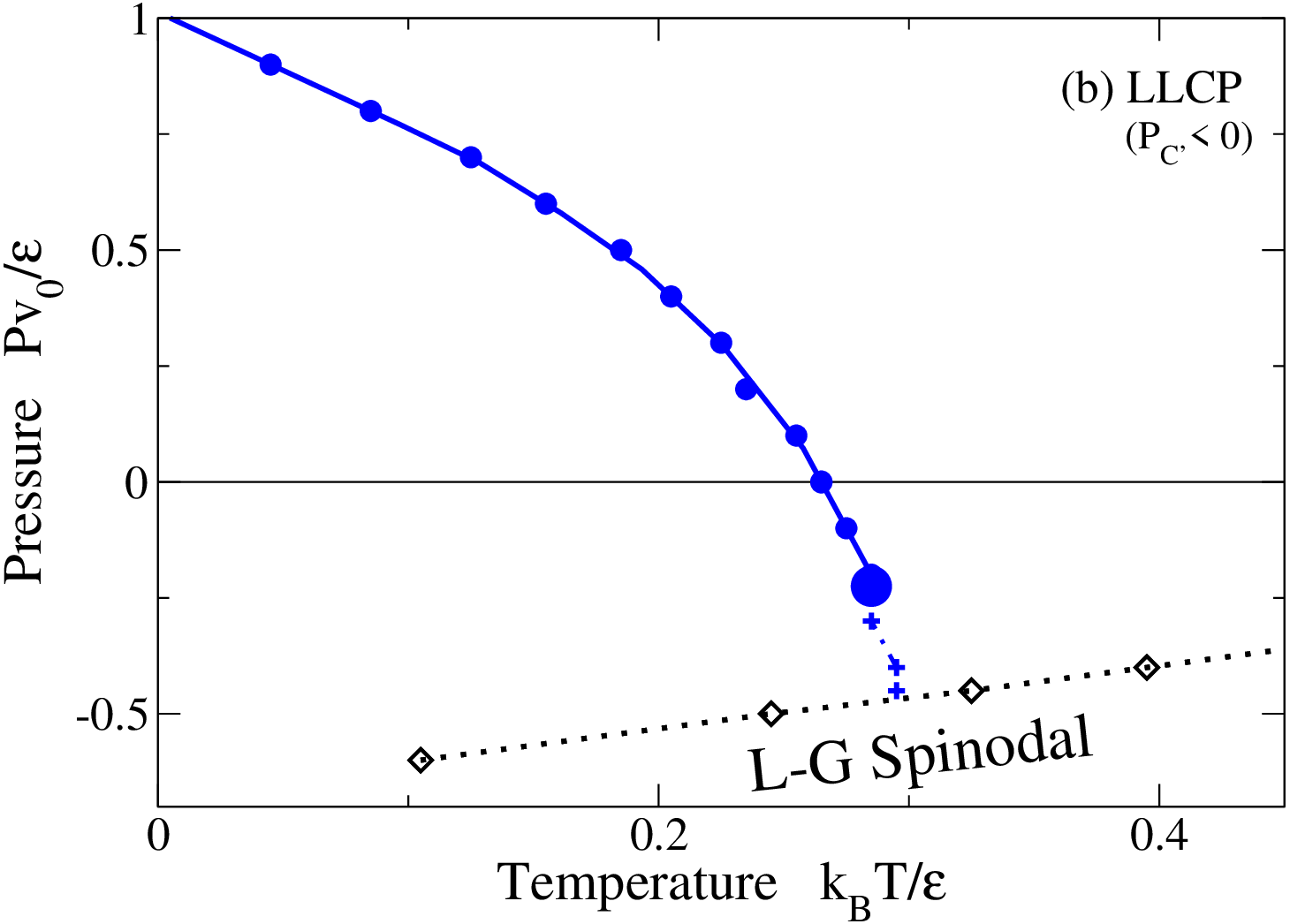}
\includegraphics[scale=0.3]{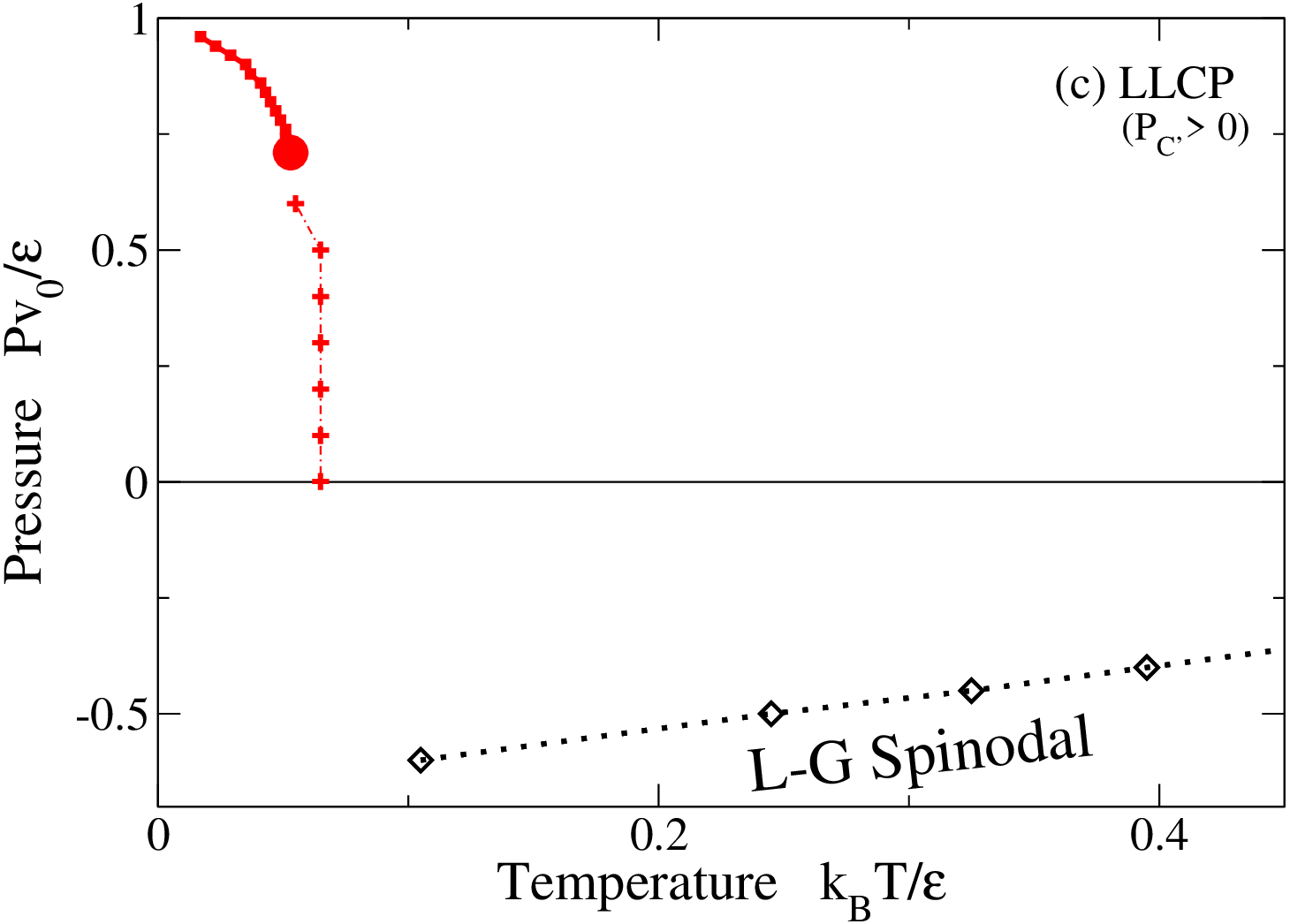}
\end{center}
\caption{Phase diagrams from MC simulations. We fix the parameters 
$\tilde{J}=0.5$, $v_{\rm HB}/v_0=0.5$ and vary the parameter
  $\tilde{J}_\sigma$. 
(a) 
For $\tilde{J}_\sigma = 0.5$ (i.e. $\tilde{J}_\sigma>a+b\tilde{J}$
where $a$ and $b$ are given in the text and both are approximately 1/3), we find 
a liquid--liquid phase transition
  (thick line with circles) ending on the superheated liquid-to-gas
spinodal (dotted line with diamonds)
as in the CPF scenario. 
(b)
  For $\tilde{J}_\sigma = 0.3$ (i.e. $\tilde{J}_\sigma> \tilde{J}/2$), the
  liquid--liquid phase transition 
ends in a liquid--liquid critical point
(LLCP) at negative  pressure.
(c) 
  For $\tilde{J}_\sigma = 0.05$ (i.e. $\tilde{J}_\sigma < \tilde{J}$), the LLCP ends
  at positive 
  pressure and the line of  specific heat maxima
(crosses), emanating from the LLCP, is shown only for positive pressure.
Errors are of the order of the
  symbol sizes. Lines are guides for the eyes.
Other model parameters are as for MF
  calculations (see text).}
\label{MC}
\end{figure}

We simulate this system for
$N=10^4$ molecules arranged on a square lattice, adopting Wolff's
algorithm to equilibrate at low $T$ \cite{mazza}, for different values
of $\tilde{J_\sigma}$, keeping constant $\tilde{J}=0.5$,
and $v_{\rm HB}/v_0=0.5$ (Fig.~\ref{MC}).

For large values of $\tilde{J_\sigma}$ ($\tilde{J_\sigma}=0.5> a+b
\tilde{J}$), we find a 
HDL--LDL first--order phase transition that merges with the
superheated liquid spinodal as in the CPF scenario (Fig.~\ref{MC}a).  
At lower
$\tilde{J}_\sigma$ ($\tilde{J_\sigma}=0.3> \tilde{J}/2$), a HDL--LDL critical point
appears at $P<0$, from which emanates the locus of $C_P$ maxima
(used here as an approximation of the liquid--liquid Widom line),
which intersects the 
superheated liquid spinodal (Fig.~\ref{MC}b).  
By further decreasing $\tilde{J_\sigma}$
($\tilde{J_\sigma}=0.05<\tilde{J}/2$), the HDL--LDL critical point occurs at
$P>0$, with the line of $C_P$ maxima
intersecting the $P=0$ axis (Fig.~\ref{MC}c).  
For $\tilde{J_\sigma}=0.02$, approaching zero, we find
that the temperature of the HDL--LDL critical point approaches zero
and the critical pressure increases toward the value $P= \epsilon/v_0$
independent of $\tilde{J}_\sigma$. In this case, we can show that Eq.~(\ref{PC'})
still holds, but with $\tilde{J}^*\equiv \tilde{J}$.
The line of $C_P$ maxima approaches
the $T=0$ axis for $\tilde{J}_\sigma \rightarrow 0$.  These results
confirm the qualitative behavior
found with the MF calculations.

\section{Comparison with other thermodynamic models}

To show that our analysis offers a general framework within which to analyze 
the supercooled water phase diagram in terms of the
interplay between the strengths of the directional contribution to the
H bond interaction and its cooperative part, we compare
our results with those from other thermodynamic models 
that can reproduce more than one scenario by tuning appropriate parameters
\cite{psgsa,bds,tanaka2000}.

One free energy model with cooperative interactions is the one
introduced by Tanaka \cite{tanaka2000}. He shows that, as in
the SF scenario, water's anomalies are the effect of 
the excitation of locally favored structures upon cooling, which have 
lower energy and larger volume than normal-liquid structures.
As in our model, in Tanaka's model increasing the cooperativity 
among excitations of locally favored structures leads to the LLCP scenario.
Moreover, Tanaka's model LLCP is regulated by relations such as our
Eq.(\ref{TC'}) and (\ref{PC'}). Therefore, by increasing the strength
of the cooperative interaction, the LLCP will eventually reach the
limit of stability of the liquid, as in the CPF/SL scenario.

We next consider the free energy model 
introduced by Poole et al. \cite{psgsa}, in which a van der Waals
free energy is augmented to include the effect of H bond formation.
The H bond interaction is characterized by two free parameters:
the strength of the H bond, and a geometrical constraint on H bond formation.
The fraction of molecules that form H bonds with decreased energy
and entropy is determined by a distribution over molar volumes, the
width of which is $\Delta$.
Poole et al. show that, by keeping $\Delta$ fixed, their model
displays a SL scenario for weak H bond energy, and a LLCP
at positive pressure for strong H bond energy.
This corresponds in our model to increase the H bond coupling $\tilde{J}$ from
$\tilde{J}<(\tilde{J_\sigma}-a)/b$ to $\tilde{J}>2\tilde{J_\sigma}$, 
while keeping $\tilde{J_\sigma}>a$ fixed. 


Next we study the effect of varying the other H bond parameter in
the Poole et al. model, the width $\Delta$.  Keeping the H bond
energy fixed, we produce the LLCP phase behavior at large
$\Delta$ and the SL phase behavior at small $\Delta$.  Hence a
decrease of $\Delta$ has the same 
effect on the phase diagram as an increase in the H bond cooperativity
in our model.  

This result is consistent with that of Borick
et al. ~\cite{bds} for their Hamiltonian model 
that incorporates the cooperativity of H bonds trough the same
mechanism used by Poole et al., i.e. by adopting a distribution with
width $\Delta$ that makes the H bond strength density dependent. By 
decreasing $\Delta$,
Borick et al. find
that the LLCP moves to lower $P$ and higher $T$. 
This behavior makes sense physically, as a more all-or-nothing
distribution of H bonds (small $\Delta$) implies a more cooperative process of
bond formation.  It also implies that the models of Poole et al. and
Borick et al. give rise to the SF scenario only in the limiting case
of infinite $\Delta$.

We conclude that all four models give a consistent physical picture.
This suggests that our result, expressed in terms of strength of the directional and
cooperative components of the H bond, as summarized in Fig.~\ref{2},
is general.


\section{Estimates from experimental data}

In the framework of the scheme presented here, in which directionality
and cooperativity are the two relvant physical parameters,
we propose that the way to understand which scenario best describes water
is to probe the energy of the covalent part of the H bond interaction~\cite{hbond0} and the energy of the
cooperative component of the H bond interaction~\cite{hbond1,hbond2,hbcoop}.
Experiments measure H bonds in ice Ih to be approximately 3 kJ/mol stronger than in liquid water~\cite{newref}.
Attributing this increase to a cooperative interaction among H bonds~\cite{energy-coop},
we can estimate the value of $J_\sigma$ in the cell model to be $\approx$ 1.0 kJ/mol.
An estimate of the van der Waals attraction, based on isoelectronic molecules at optimal separation, yields $\epsilon \approx$ 5.5 kJ/mol~\cite{energy-eps}.  The optimal H bond energy, $E_{\rm HB}$, has been measured to be $\approx$23.3 kJ/mol~\cite{suresh}.  By considering tetrahedral clusters of H bonded molecules, with H bond and van der Waals interactions between n.n. molecules (and appropriately reduced van der Waals interactions between second and third n.n. molecules), we derive the value for the directional component of the H bond, $J\approx$ 12.0 kJ/mol.
Other experimental estimates suggest that breaking the directional
component of the H bond requires $\approx$ 6.3 kJ/mol~\cite{newref2}.

Both estimates from experiments 
fall within the range of $1.1\leq \tilde{J}\leq
2.2$, with $\tilde{J}_\sigma\simeq 0.2$, i. e. with
$\tilde{J}_\sigma< \tilde{J}/2$.  Therefore, within our model,
these values 
lead to the LLCP scenario with $P_{C'}>0$. In particular,
MF calculations with $\tilde{J}_\sigma= 0.2$, $\tilde{J}= 2.2$ and
$v_{\rm HB}=2 v_0$, predict a LLCP at $T_{C'}=0.25
\epsilon/k_B$ and $P_{C'}=3.5 \epsilon/(v_0 k_B)$. 

\section{Conclusions}

We have shown that a microscopic cell model of water, by
taking into account the cooperativity among H bonds, is able
to produce phase behaviors consistent with any of the proposed
scenarios for water's phase diagram.  It is the amount of
cooperativity in relation to the strength of the directional component
of the H bond that establishes
 which scenario holds.  For no amount of
cooperativity, the SF scenario is recovered.  By increasing the
amount of cooperativity in relation to the H bond directional strength, a liquid--liquid
transition grows out from the $T=0$ axis, ending in a LLCP.  With
sufficiently strong cooperativity, this LLCP lies beyond the region
of stable liquid states, leaving only the liquid--liquid transition,
consistent with the CPF scenario.  In this case the spinodal associated
with the transition acts as the line predicted in the SL scenario.


Comparison with previous models gives consistent results.
Hence we argue that each of the four scenarios proposed for the phase
diagram of liquid water may be viewed as a special case of our
general scheme. This scheme is based on the assumption
that water-water interaction is characterized by an isotropic
component, a directional component and a cooperative component, and
that H bond formation leads to an open local structure. Alternative
mechanisms, based only on isotropic interactions~\cite{fmsbs,jagla,xu,f,deOFNB} or only on
directional interactions~\cite{starr} have been considered and their
relevance for the water case is an open question.
Finally, estimates for the three components of the H bond interaction, based on experimental data,
lead to the conclusion that the LLCP scenario with a positive critical pressure holds for water.

\begin{acknowledgments}
 We thank P. Poole, S. Sastry, F. Sciortino, and F. Starr for helpful
 discussions, and NSF grant CHE0616489 for support. G.F. thanks 
the Spanish Ministerio de Ciencia e Innovaci\'on grant
 FIS2009-10210 (co-financed FEDER). 

\end{acknowledgments}

\end{document}